\documentclass{elsarticle}
\usepackage{amsmath, amssymb}
\usepackage[small,justification=centering]{caption}
\usepackage{graphicx,dblfloatfix}
\usepackage{geometry}
\usepackage{hyperref}
\usepackage{mathtools}
\usepackage{physics}
\usepackage{relsize}
\usepackage{subcaption}
\usepackage{color}
\usepackage[inline]{enumitem}

\graphicspath{{./figs/}{.}}

\biboptions{numbers,sort&compress}

\newcommand*\widebar[1]{%
	\hbox{%
		\vbox{%
			\hrule height 0.5pt 
			\kern0.5ex
			\hbox{%
				\kern-0.2em
				\ensuremath{#1}%
				\kern-0.0em
			}%
		}%
	}%
}

\begin{document}
\title{A large eddy lattice Boltzmann simulation of magnetohydrodynamic turbulence}
\author{Christopher Flint}
\author{George Vahala}
\address{Department of Physics, William \& Mary, Williamsburg, Virginia 23185}
\date{\today}

\begin{abstract}
 
  Large eddy simulations (LES) of a lattice Boltzmann magnetohydrodynamic (LB-MHD) model are performed for the unstable magnetized Kelvin-Helmholtz jet instability.  This algorithm is an extension of Ansumali \textit{et al.} \cite{Ansumali} to MHD in which one performs first an expansion in the filter width on the kinetic equations followed by the usual low Knudsen number expansion.  These two perturbation operations do not commute.  Closure is achieved by invoking the physical constraint that subgrid effects occur at transport time scales.  The simulations are in very good agreement with direct numerical simulations.  
\end{abstract}
\maketitle
\section{Introduction}\label{intro}

   Recently \cite{flintLEStheory} we derived a first-principles two dimensional (2D) magnetohydrodynamic (MHD) large eddy simulation (LES) model based on first filtering the lattice Boltzmann (LB) representation of MHD \cite{DellarBulkViscosity,DellarMHDLB,DellarMRT,DellarMoments,dellar2013resistivity,dellar2014cubicDefects,flint9bitMHD,vahala2008mhd} after which one applies the Chapman-Enskog limits to recover the final LES-MHD fluid equations. In essence, we extended to MHD the 2D Navier-Stokes (NS) LES-LB model of Ansumali \textit{et. al.} \cite{Ansumali} who exploited the non-commutativity of these two operations.  (Of course, if one first applied the Chapman-Enskog limit to LB and then filtering one would land in the conventional quagmire of an LES closure problem.)  A technical difficulty with the Ansumali \textit{et. al.} model is that in 2D NS there is an inverse energy cascade to large spatial scales thereby rendering subgrid modeling non-essential.  In 2D MHD, however, the energy cascades to small spatial scales as in 3D - and so makes it attractive to perform the LES-LB-MHD simulations in which there can be a substantial amount of excited subgrid modes.  Here we present some preliminary LES-LB-MHD simulations of our model and compare the results with some direct numerical simulations (DNS).  As Ansumali  \textit{et. al.} \cite{Ansumali} did not perform any simulations on their LES-LB-NS model, these are the first such LES-LB-MHD simulations when one has first filtered the underlying LB representation, followed by the conventional small Knudsen number expansion.  
  
   The backbone of any LES \cite{Carati2001,Carati2002,Clark,Vreman,EarlySmagMHD1,EarlySmagMHD2,EarlySmagMHD3,EarlySmagMHD4,Pope,premnath2009,ChenDoolen,Ansumali9,Ansumali11,Ansumali12,Ansumali13,Ansumali14,krafczyk2003large,chen2009simple,girimajiDNSLES,flintLEStheory} is the introduction of a spatial filter function to smooth out field fluctuations on the order of the filter width $\Delta$.  Thus for the mean velocity 
\begin{equation}\label{VelFilter}
\mathbf{\bar u}\left(\vec{r},\Delta\right)=\int_{-\infty}^\infty \mathbf{u} \left(\vec{r}\,'-\vec{r}\,\right)G\left(\vec{r}\,',\Delta\right)d\vec{r}\,'  .
\end{equation}
In general, the filtering results in the standard closure problem.  Previous LB-LES-NS modeling \cite{chen2009simple,krafczyk2003large} have first considered the Chapman-Enskog expansion followed by filter and thus have concentrated on the Smagorinsky closure for the subgrid stresses.  It has been pointed out \cite{girimajiDNSLES} that in the conventional NS-LES closure, the subgrid stresses are assumed to be in equilibrium with the filtered strain. However, in LB-LES-NS the stresses relax towards the filtered strain at a rate dictated by the current eddy-viscosity, thereby permitting some spatio-temporal memory effects that are absent when applying LES directly to the continuum equations.  In essence this gives an edge to any LB - approach.

In the Ansumali approach, however,  one first performs a perturbation expansion in the filter width $\Delta$ followed by the standard LB Chapman-Enskog expansion in the Knudsen number $\mathrm{Kn}$.  These two perturbation expansions do not commute.  Closure is now achieved by making the physically plausible assumption that eddy transport effects occur at the transport time scale and this results in the scaling $\Delta \simeq \order{\mathrm{Kn}^{1/2}}$.  One still retains the LB effects of spatial -temporal memory as noted earlier \cite{chen2009simple,krafczyk2003large} .

\section{LES-LB-MHD Model}
For completeness we briefly review the essentials of our LES-LB-MHD model \cite{flintLEStheory} that yields the following closed set of filtered MHD equations (without further approximations)
for the filtered density $\left( \bar \rho \right)$, momentum $ \left(\, \widebar {\rho \mathbf u} \right) $ and the magnetic field $(\overline{ \mathbf B})$ in multiple relaxation time (MRT)

\begin{footnotesize}
\begin{subequations}\label{FinalLESeq}
\begin{align}\label{Density}
&\partial_t \, {\widebar{\rho}} + \nabla \cdot \, \widebar{\rho \mathbf u} = 0 \qquad , \qquad \nabla \cdot \overline{\mathbf B} = 0  
\\[10px]
\label{FinalMomentum}
&\begin{aligned}
\begin{split}
\partial_{t} \left(\, \widebar{\rho \mathbf u} \right)&+ \nabla \cdot \left( \frac{\widebar{\rho \mathbf u} \ \widebar{\rho \mathbf u}}{\bar\rho} \right) = - \nabla \bar p + \nabla \cdot \left( \overline{\mathbf B} \, \overline{\mathbf B} \right) - \frac{1}{2} \nabla \left( \overline{\mathbf B} \cdot \overline{\mathbf B} \right) + \left( \xi + \frac{1}{3} \nu \right) \nabla \left( \nabla \cdot \widebar{\rho \mathbf u} \right) + \nu \nabla^2 \, \widebar{\rho \mathbf u} \\
  & - \nabla \cdot \left\lbrace \frac{6\nu}{6\nu+1} \frac{\Delta^2}{12\bar\rho} \left[ \left( \partial_\beta \left( \, \widebar{\rho \mathbf u} \right) \right) \left( \partial_\beta \left( \, \widebar{\rho \mathbf u} \right) \right) -  \frac{\partial_\beta \bar p}{\bar p} \left( \widebar{\rho \mathbf u} \left( \partial_\beta \left( \, \widebar{\rho \mathbf u} \right) \right) + \left( \partial_\beta \left( \, \widebar{\rho \mathbf u} \right) \right) \widebar{\rho \mathbf u} -  \widebar{\rho \mathbf u} \ \widebar{\rho \mathbf u} \frac{\partial_\beta \bar p}{\bar p} \right) \right] \right\rbrace \\
&- \nabla  \left\lbrace \left( \frac{s_4}{4} + \frac{s_7}{20} - \frac{3 s_8}{10} \right) \frac{\Delta^2}{12\bar\rho} \left[ \left( \partial_\beta \left( \, \widebar{\rho \mathbf u} \right) \right) \cdot \left( \partial_\beta \left( \, \widebar{\rho \mathbf u} \right) \right) - \frac{\partial_\beta \bar p}{\bar p} \left( 2 \, \widebar{\rho \mathbf u} \cdot \left( \partial_\beta \left( \, \widebar{\rho \mathbf u} \right) \right) -  \widebar{\rho \mathbf u} \cdot \widebar{\rho \mathbf u} \frac{\partial_\beta \bar p}{\bar p} \right) \right] \right\rbrace\\
  & - \frac{6\nu}{6\nu+1} \frac{\Delta^2}{12} \left\lbrace \frac{1}{2} \nabla \left[ \left( \partial_\beta \overline{\mathbf B} \right) \cdot \left( \partial_\beta \overline{\mathbf B} \right) \right]  - \nabla \cdot \left[ \left( \partial_\beta \overline{\mathbf B} \right) \left( \partial_\beta \overline{\mathbf B} \right) \right]
\right\rbrace  ,
\end{split}
\end{aligned}
\\[10px]
\label{FinalMagnetic}
&\begin{aligned}
\begin{split}
\partial_t \, \overline {\mathbf B} &= \nabla \cross \left( \frac{ \widebar{\rho \mathbf u} \cross \overline {\mathbf B}}{\bar\rho} \right) + \eta\nabla^2\,\overline {\mathbf B} + \nabla \cross \left[ \frac{\Delta^2}{12\bar\rho} \frac{6\eta}{6\eta+1} \left\lbrace \left( \partial_\beta \left( \, \widebar{\rho \mathbf u} \right) \right) \cross \left( \partial_\beta \, \overline {\mathbf B} \right) \right. \right. \\
   &\left. \left. -  \frac{\partial_\beta \bar p}{\bar p} \left( \left( \, \widebar{\rho \mathbf u} \right) \cross \left( \partial_\beta \, \overline {\mathbf B} \right) + \left( \partial_\beta \left( \, \widebar{\rho \mathbf u} \right) \right) \cross \overline {\mathbf B} -  \frac{ \left( \partial_\beta \bar p \right) }{\bar p} \left( \, \widebar{\rho \mathbf u} \right) \cross \overline {\mathbf B} \right) \right\rbrace \right] .
\end{split}
\end{aligned}
\end{align}
\end{subequations}
\end{footnotesize}

\noindent where $s_3 \ldots s_8$ are relaxation rates, and in this isothermal model, the pressure is directly related to the density  $p = \rho  c_s^2 = \frac{\rho}{3}$, in lattice units ($c_s$ is the sound speed).  The transport coefficients (shear viscosity $\nu$, bulk viscosity $\xi$ and resistivity $\eta$) are determined from the LB-MRT for the particle distribution function (the $s's$) and the single magnetic distribution function relaxation rate $s_m$:
\begin{equation}
\nu =\frac{1}{3s_3}-\frac{1}{6}=\frac{1}{3s_4}-\frac{1}{6}
\end{equation}
\begin{equation}
\xi =-\frac{1}{9}-\frac{1}{9s_4}-\frac{1}{15s_7}+\frac{2}{5s_8}
\end{equation}
\begin{equation}
\eta =\frac{1}{3s_{m}}-\frac{1}{6}
\end{equation}


We now summarize our computational LB-LES-MHD model that underlies Eqs. \eqref{FinalLESeq}.   For 2D MHD, we consider an LB model with 9-bit lattice 
\begin{eqnarray}
& \label{LBKinEqn}\left( \partial_t + \partial_\gamma c_{\gamma i} \right) f_i = \sum_j s^{'}_{ij} \left( f_j^{(\mathrm{eq})} - f_j \right)  , \quad  i = 0 ... 8  \\
& \label{LBMagEqn}\left( \partial_t + \partial_\gamma c_{\gamma i} \right) \vec g_i = s^{'}_{m} \left( \vec g_i^{\,(\mathrm{eq})} - \vec g_i \right)   ,  \quad i = 0 ... 8
\end{eqnarray}
with the moments $\sum_i f_i = \rho$,  $\sum_i f_i  \vec c_i = \rho \vec u$, and $\sum_k \vec g_k = \vec B$.  Here the summation convention is employed on the vector nature of the fields (using Greek indices) while for Roman indices, correspond to the corresponding lattice vectors for the kinetic velocities $\vec c_i$, there is no implicitly implied summation.  The lattice is just the axes and diagonals of a square (along with the rest particle $i = 0$).  $s^{'}_{ij}$ are the MRT collisional relaxation rate tensor for the $f_i$ while the SRT $s^{'}_{m}$ is the collisional relaxation rate for $\vec g_i$.  These kinetic relaxation rates determine the MHD viscosity and resistivity transport coefficients.  (Of course, more sophisticated LB models can be formed by MRT on the $\vec g_k$ equations, but for this first reported LB-LES-MHD simulation we will restrict ourselves to the simpler SRT model)

A convenient choice of the relaxation distribution functions, will under Chapman-Enskog, yield the MHD equations
\begin{gather}
\label{feq}f_i^{(\mathrm{eq})} = w_i \rho \left[ 1 + 3\left( \vec c_i \cdot \vec u \right) + \frac{9}{2}\left( \vec c_i \cdot \vec u \right)^2 - \frac{3}{2} \vec u^{\,2} \right] + \frac{9}{2} w_i \left[ \frac{1}{2} \vec B^2 \vec c_i^{\,2} - \left( \vec B \cdot \vec c_i \right)^2 \right]  , i = 0, .. ,8 \\
\label{geq}\vec g_i^{\,(\mathrm{eq})} = w_i^{'} \!\left[ \vec B + 3 \left\lbrace \left( \vec c_i \cdot \vec u \right) \vec B - \left( \vec c_i \cdot \vec B \right) \vec u \right\rbrace \right] , i = 0, .., 8
\end{gather}
where the $w's$ are appropriate lattice weights.  In the operator-splitting solution method of collide-stream, it is most convenient to perform the collision step in moment-space (because of collisional invariants of the zeroth and 1st moment of $f_i$, and the zeroth moment of $\vec g_i$.), while the streaming is optimally done in the $( f_i, \vec g_i)$-space.   Moment space $(M_i , \vec N_i)$ is defined by
 \begin{equation}\label{Transformation}
\begin{array}{ccc}
M_i = \sum_{j=0}^8 \mathrm T_{ij}f_j &, & \vec N_{i} = \sum_{q=0}^8  \mathrm T_{\mathrm m,iq} \vec g_{q}
\end{array}
\end{equation}
with the 1-1  constant transformation matrices, $\mathrm T$ and $\mathrm T_{\mathrm m}$ given by
\begin{eqnarray}
& \label{FluidTMatrix}
\mathrm T=\left( \def\arraystretch{1.2}\begin{array}{c}
\boldsymbol{1} \\ 
c_x \\ 
c_y \\ 
c_xc_y \\ 
c^2_x-c^2_y \\ 
3c_xc^2_y-2c_x \\ 
3c_yc^2_x-2c_y \\ 
4\cdot \boldsymbol{1}-9\left(c^2_x+c^2_y-2c^2_xc^2_y\right) \\ 
4\cdot \boldsymbol{1}-4\left(c^2_x+c^2_y\right)+3c^2_xc^2_y \end{array}
\right)=\left( \def\arraystretch{1.2}\begin{array}{rrrrrrrrr}
1 & 1 & 1 & 1 & 1 & 1 & 1 & 1 & 1 \\ 
0 & 1 & 0 & -1 & 0 & 1 & -1 & -1 & 1 \\ 
0 & 0 & 1 & 0 & -1 & 1 & 1 & -1 & -1 \\ 
0 & 0 & 0 & 0 & 0 & 1 & -1 & 1 & -1 \\ 
0 & 1 & -1 & 1 & -1 & 0 & 0 & 0 & 0 \\ 
0 & -2 & 0 & 2 & 0 & 1 & -1 & -1 & 1 \\ 
0 & 0 & -2 & 0 & 2 & 1 & 1 & -1 & -1 \\ 
4 & -5 & -5 & -5 & -5 & 4 & 4 & 4 & 4 \\ 
4 & 0 & 0 & 0 & 0 & -1 & -1 & -1 & -1 \end{array}
\right) \\
\text{and}
& \label{MagTMatrix}
\mathrm {T_m}=\left( \def\arraystretch{1.2} \begin{array}{c}
\boldsymbol{1} \\ 
c_x \\ 
c_y \\ 
c_xc_y \\ 
c^2_x \\ 
c^2_y \\ 
c_x^2 c_y \\ 
c_x c_y^2\\ 
c_x^2 c_y^2 \end{array}
\right)=\left( \def\arraystretch{1.2} \begin{array}{rrrrrrrrr}
1 & \phantom{-}1 & \phantom{-}1 & 1 & 1 & \phantom{-}1 & 1 & 1 & 1 \\ 
0 & 1 & 0 & -1 & 0 & 1 & -1 & -1 & 1 \\ 
0 & 0 & 1 & 0 & -1 & 1 & 1 & -1 & -1 \\ 
0 & 0 & 0 & 0 & 0 & 1 & -1 & 1 & -1 \\ 
0 & 1 & 0 & 1 & 0 & 1 & 1 & 1 & 1 \\ 
0 & 0 & 1 & 0 & 1 & 1 & 1 & 1 & 1 \\ 
0 & 0 & 0 & 0 & 0 & 1 & 1 & -1 & -1 \\ 
0 & 0 & 0 & 0 & 0 & 1 & -1 & -1 & 1 \\ 
0 & 0 & 0 & 0 & 0 & 1 & 1 & 1 & 1 \end{array}
\right)
\end{eqnarray}          
The $x$ and $y$ components of the $9$-dimensional lattice vectors are
\begin{equation}
\begin{array}{ccc}
c_x=\left\{0,1,0,-1,0,1,-1,-1,1\right\} &, &c_y=\{0,0,1,0,-1,1,1,-1,-1\}
\end{array}
\end{equation}                                

In terms of conserved moments, we can write
\begin{gather}\label{KinMomentEq}
\arraycolsep=5pt
\begin{array}{lll}
M_0^{(eq)} = M_0 = \rho & 
M_3^{(eq)} = \frac{\rho u_x \rho u_y}{\rho} - B_x B_y &
M_6^{(eq)} = -\rho u_y \\
M_1^{(eq)} = M_1 = \rho u_x &
M_4^{(eq)} = \frac{\left( \rho u_x \right)^2 - \left( \rho u_y \right)^2}{\rho} - B_x^2 + B_y^2 &
M_7^{(eq)} = -3\frac{\left( \rho u_x \right)^2 + \left( \rho u_y \right)^2}{\rho} \\
M_2^{(eq)} = M_2 = \rho u_y & 
M_5^{(eq)} = -\rho u_x &
M_8^{(eq)} = \frac{5}{3}\rho - 3\frac{\left( \rho u_x \right)^2 + \left( \rho u_y \right)^2}{\rho} \\
\end{array}
\\[2ex]
\label{MagMomentEq}
\arraycolsep=20pt
\begin{array}{lll}
N_{\alpha 0}^{(eq)} = N_{\alpha 0} = B_\alpha & 
N_{\alpha 3}^{(eq)} = 0 &
N_{\alpha 6}^{(eq)} = \frac{1}{3}\left(\rho u_y B_\alpha - \rho u_\alpha B_y\right) 
\\[1ex]
N_{\alpha 1}^{(eq)} = \rho u_x B_\alpha - \rho u_\alpha B_x &
N_{\alpha 4}^{(0)} = \frac{B_\alpha}{3} &
N_{\alpha 7}^{(eq)} = \frac{1}{3}\left(\rho u_x B_\alpha - \rho u_\alpha B_x\right)
\\[1ex]
N_{\alpha 2}^{(eq)} = \rho u_y B_\alpha - \rho u_\alpha B_y &
N_{\alpha 5}^{(0)} = \frac{B_\alpha}{3} &
N_{\alpha 8}^{(eq)} = \frac{B_\alpha}{9}
\end{array}
\end{gather}

\subsection{Filtering LB}
In applying filtering to the LB Eqs. \eqref{LBKinEqn} and \eqref{LBMagEqn}, only the nonlinear terms in the relaxation distributions, Eqs. \eqref{feq} and \eqref{geq}, require further attention.   On applying perturbations in the filter width $\Delta$ we immediately see that
\begin{equation}\label{FilterExampleSmall}
\overline{\left( XY \right) } = \widebar{X} \, \widebar{Y} + \frac{\Delta^{2}}{12} \left( \partial_{\beta} \, \widebar{X} \right) \left( \partial_{\beta} \, \widebar{Y} \right) + O\left( \Delta ^{4} \right)
\end{equation}
and 
\begin{small}\begin{equation}\label{FilterExampleLarge}
	\overline{\left( \frac{XY}{Z} \right) } = \,  \frac{\widebar{X} \, \widebar{Y}}{\widebar{Z}} + \frac{\Delta^{2}}{12 \, \widebar{Z}} \left[ \left( \partial_{\beta} \, \widebar{X} \right) \left( \partial_{\beta} \, \widebar{Y} \right) - \frac{\left( \partial_{\beta} \, \widebar{Z} \right)}{\widebar{Z}} \left( \widebar{X} \left( \partial_{\beta} \, \widebar{Y} \right) + \, \widebar{Y} \left( \partial_{\beta} \, \widebar{X} \right) - \frac{\widebar{X} \, \widebar{Y} \left( \partial_{\beta} \, \widebar{Z} \right) }{\widebar{Z}} \right) \right] + O\left( \Delta ^{4} \right)
	\end{equation}\end{small}

\noindent for arbitrary fields $X$, $Y$, and $Z$. Moreover since collisions are performed in moment space, we need first to transform from $f^{(\mathrm{eq})}, \vec g^{(\mathrm{eq})}$ to $M^{(\mathrm{eq})}, \vec N^{(\mathrm{eq})}$ and then apply filtering in terms of the filtered collisional invariants $\widebar M_0, \widebar M_1, \widebar M_2, \widebar N_{x0}, \widebar N_{y0}$
\begin{equation}
\widebar M_i^{(\mathrm{eq})} = M_i^{(\mathrm{eq})} \left( \widebar M_0, \widebar M_1, \widebar M_2, \widebar N_{x0}, \widebar N_{y0} \right) + \Delta^2 \, \widebar M_i^{(\Delta)} \quad i=0 .... 8
\end{equation}
where the $O(\Delta^2)$ term arises from the nonlinearities.  In particular for the $\widebar M_3^{(\mathrm{eq})}$-term :
\begin{gather}\begin{aligned}
\widebar M_3^{(\mathrm{eq})} &= \frac{\widebar{\rho u_x} \; \widebar{\rho u_y}}{\bar \rho} - \widebar B_x \; \widebar B_y
\\
&+ \frac{\Delta^{2}}{12 \, \bar \rho} \left[ \left( \partial_{\beta} \, \widebar{\rho u_x} \right) \left( \partial_{\beta} \, \widebar{\rho u_y} \right) - \frac{\left( \partial_{\beta} \, \bar \rho \right)}{\bar \rho} \left( \widebar{\rho u_x} \left( \partial_{\beta} \, \widebar{\rho u_y} \right) + \, \widebar{\rho u_y} \left( \partial_{\beta} \, \widebar{\rho u_x} \right) - \frac{\widebar{\rho u_x} \, \widebar{\rho u_y} \left( \partial_{\beta} \, \bar \rho \right) }{\bar \rho} \right) \right] 
\\
&- \frac{\Delta^{2}}{12} \left( \partial_{\beta} \, \widebar{B}_x \right) \left( \partial_{\beta} \, \widebar{B}_y \right) + \mathcal O\left( \Delta ^{4} \right) \ .
\end{aligned}\end{gather}
Similarly for the other filtered equilibrium moments.

\section{LES-LB-MHD Simulation}
 The filtered LB equations are now solved, with streaming performed in distribution space and collisions in moment space.  As this is the first simulation on the LB-filtered-LES approach we have made a significant number of simplifications.  We first restrict the evolution of the filtered scalar distribution function to an SRT collision operator.  In this case the relaxation rates $s_i$ are all equal so that the 3rd term in Eq. \eqref{FinalMomentum} is automatically zero.  Moreover since nearly all LB simulations are quasi-incompressible at the fluid level, we neglect (filtered) density gradients in the moment representation of the collision operator.  Thus, for example, we approximate $\widebar M_3^{(\mathrm{eq})}$ by
\begin{gather}\begin{aligned}
\widebar M_3^{(\mathrm{eq})} &= \frac{\widebar{\rho u_x} \; \widebar{\rho u_y}}{\bar \rho} - \widebar B_x \; \widebar B_y + \frac{\Delta^{2}}{12 \, \bar \rho} \left( \partial_{\beta} \, \widebar{\rho u_x} \right) \left( \partial_{\beta} \, \widebar{\rho u_y} \right) - \frac{\Delta^{2}}{12} \left( \partial_{\beta} \, \widebar{B}_x \right) \left( \partial_{\beta} \, \widebar{B}_y \right) + \mathcal O\left( \Delta ^{4} \right) \ .
\end{aligned}\end{gather}
Also, since the last term in Eq. \eqref{FinalMagnetic} is dependent on the filtered density gradient its effects at the filtered MHD level will not be significant when we code the filtered LB system.

\begin{figure}[!h]
	\centering
	\includegraphics[width=2in, height=2.3in, keepaspectratio=true]{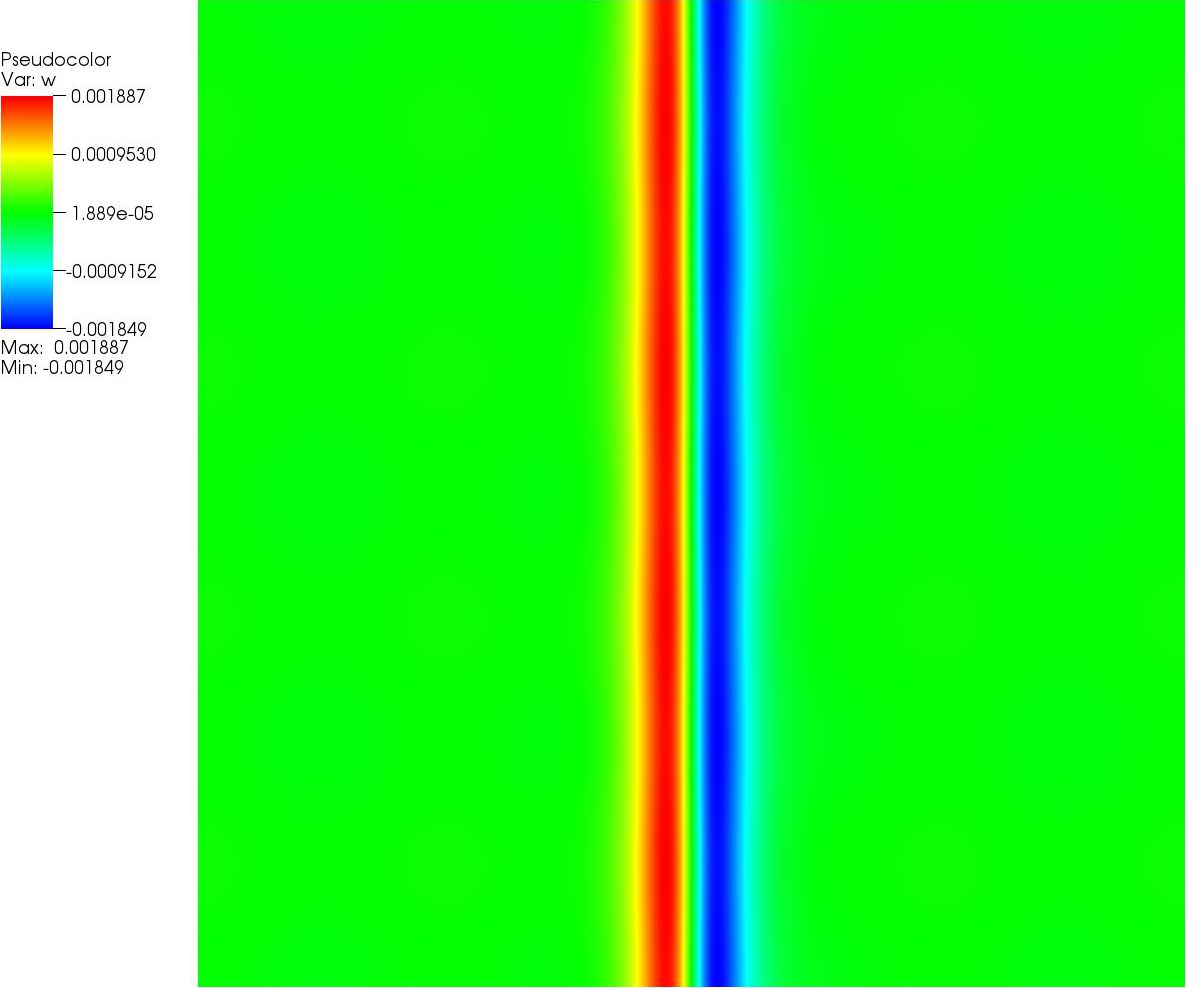}
	\caption[.]{\label{Initial}Initial vorticity profile: $U_y = U_0 \sech^2{\left(\frac{2\pi}{L}4x\right)}$ and $B_y = B_0$ where $B_0 = 0.005 U_0$.\\  Red denotes positive vorticity, while blue denotes negative vorticity.}
\end{figure}

   \begin{figure}
	\centering
	\begin{subfigure}{0.5\textwidth}
		\begin{tabular}{rc}
			\vspace{.05in}
			$t=400$k&
			\raisebox{-0.5\height}{\includegraphics[width=2in, height=2.3in, keepaspectratio=true]{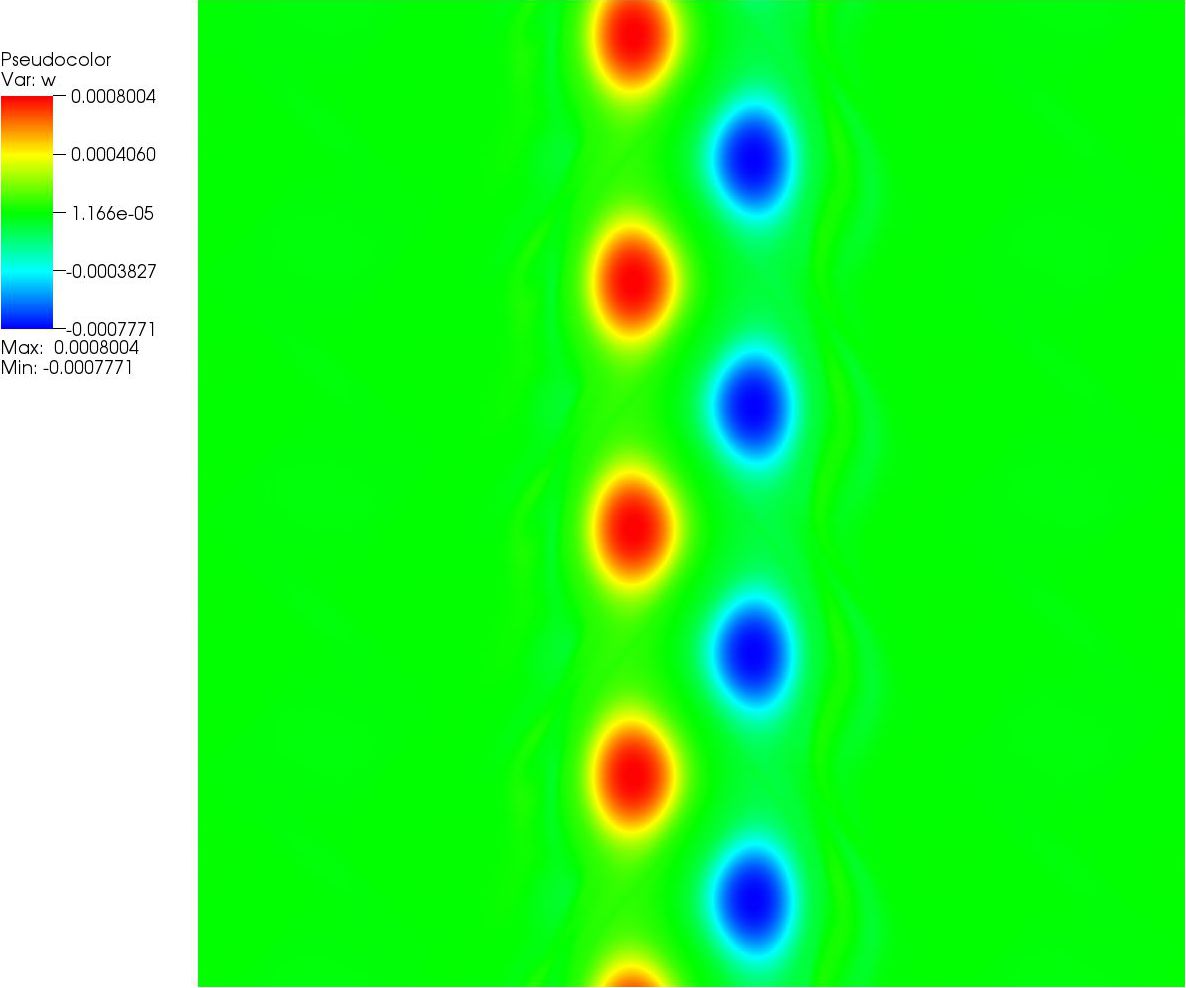}}	
			\\
			\vspace{.05in}
			$t=800$k&
			\raisebox{-0.5\height}{\includegraphics[width=2in, height=2.3in, keepaspectratio=true]{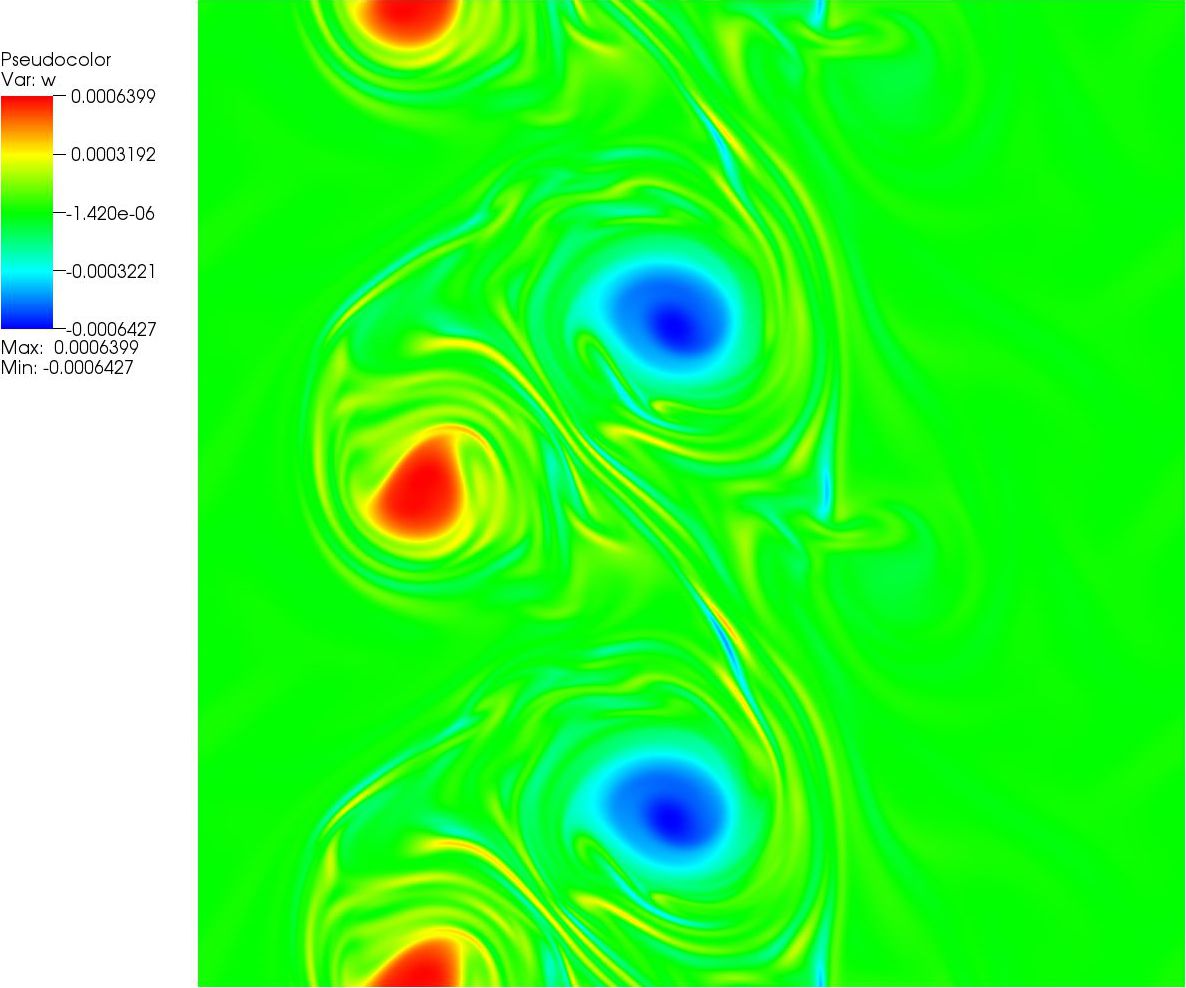}}	
			
			\\
			$t=1.56$M&
			\raisebox{-0.5\height}{\includegraphics[width=2in, height=2.3in, keepaspectratio=true]{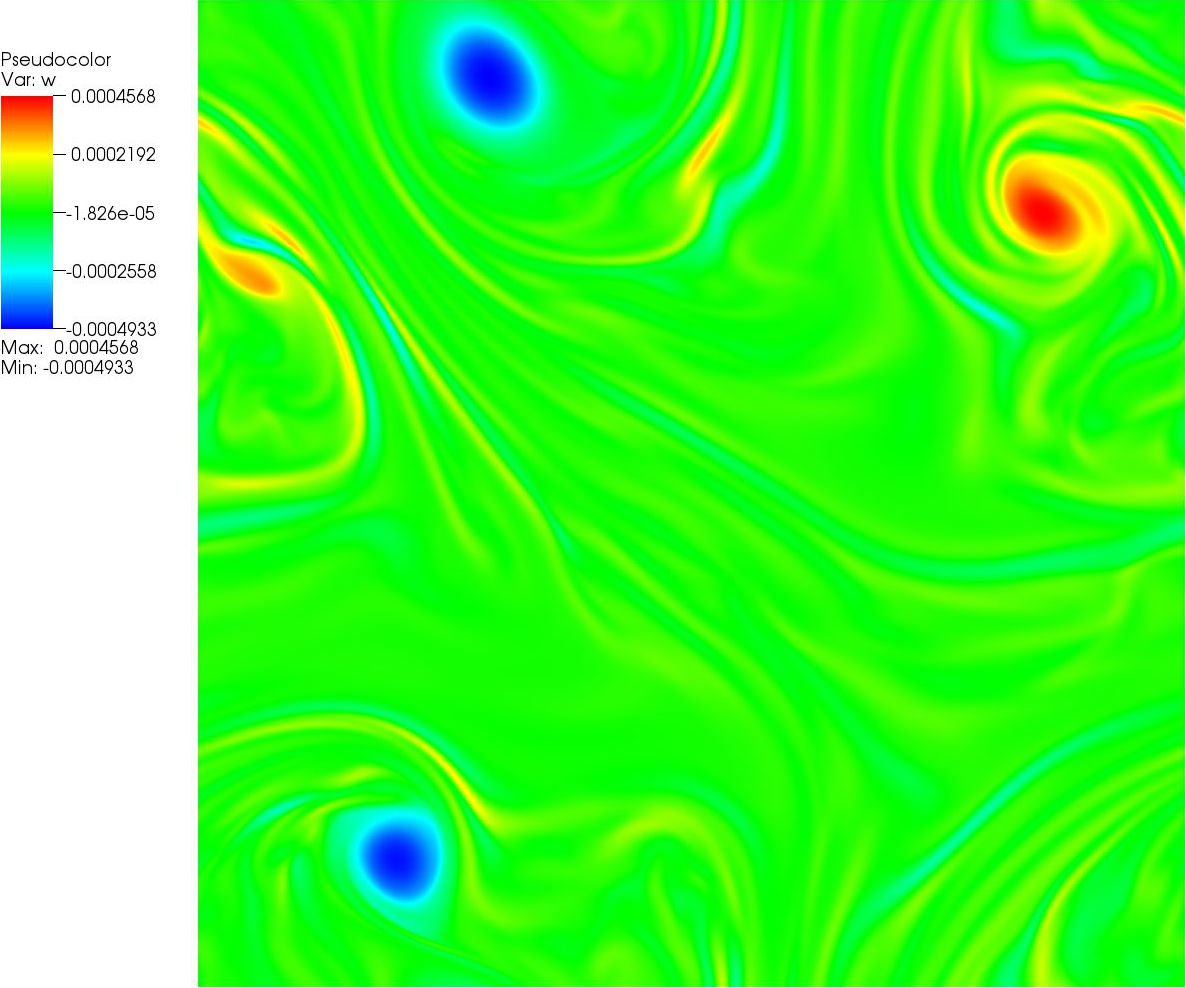}}	
		\end{tabular}
		\caption{DNS ($2048^2$)}
	\end{subfigure}%
	\begin{subfigure}{0.5\textwidth}
		\begin{tabular}{cl}
			
			\raisebox{-0.5\height}{\includegraphics[width=2in, height=2.3in, keepaspectratio=true]{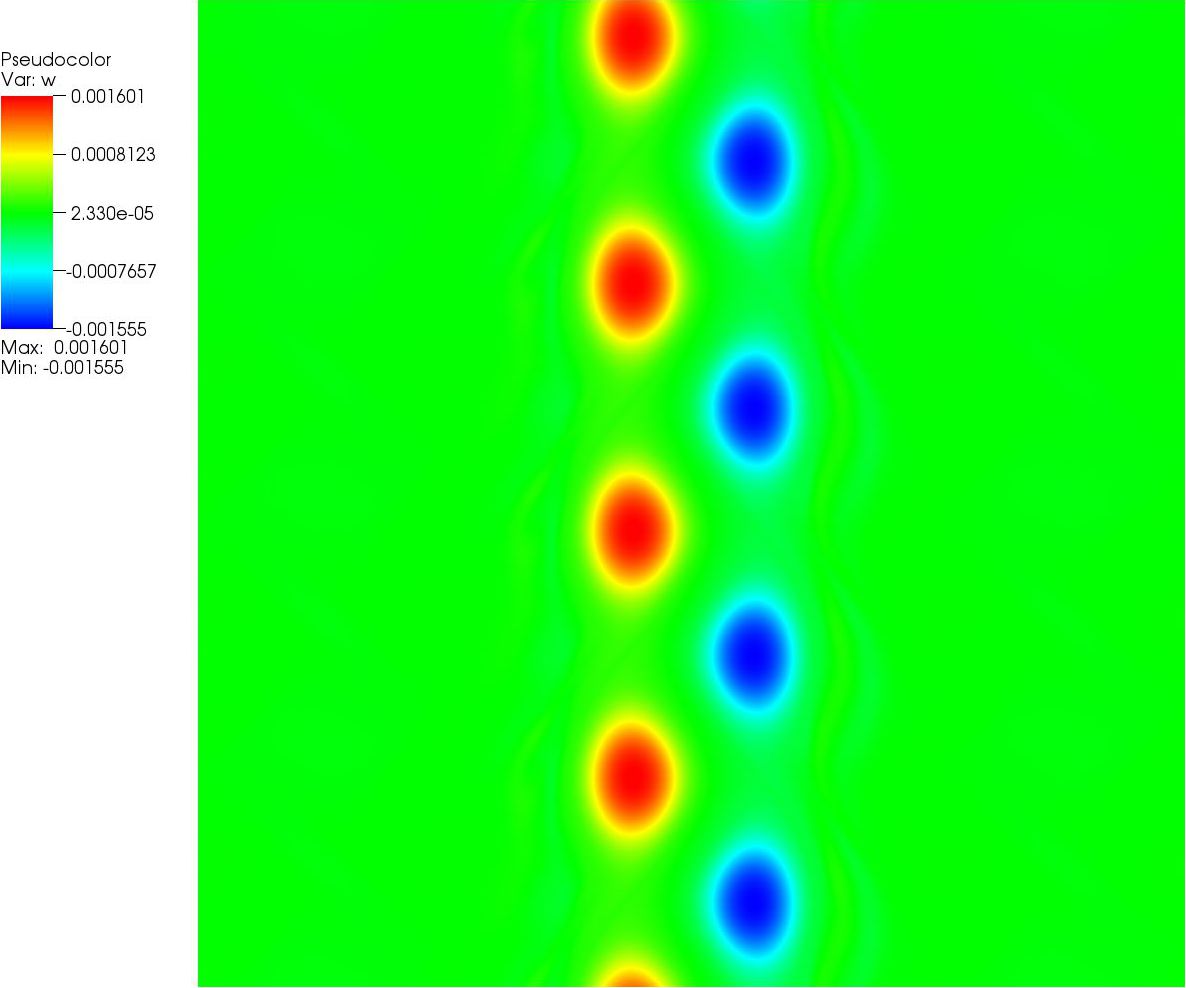}}&
			$t=200$k
			\vspace{.05in}
			\\
			
			\raisebox{-0.5\height}{\includegraphics[width=2in, height=2.3in, keepaspectratio=true]{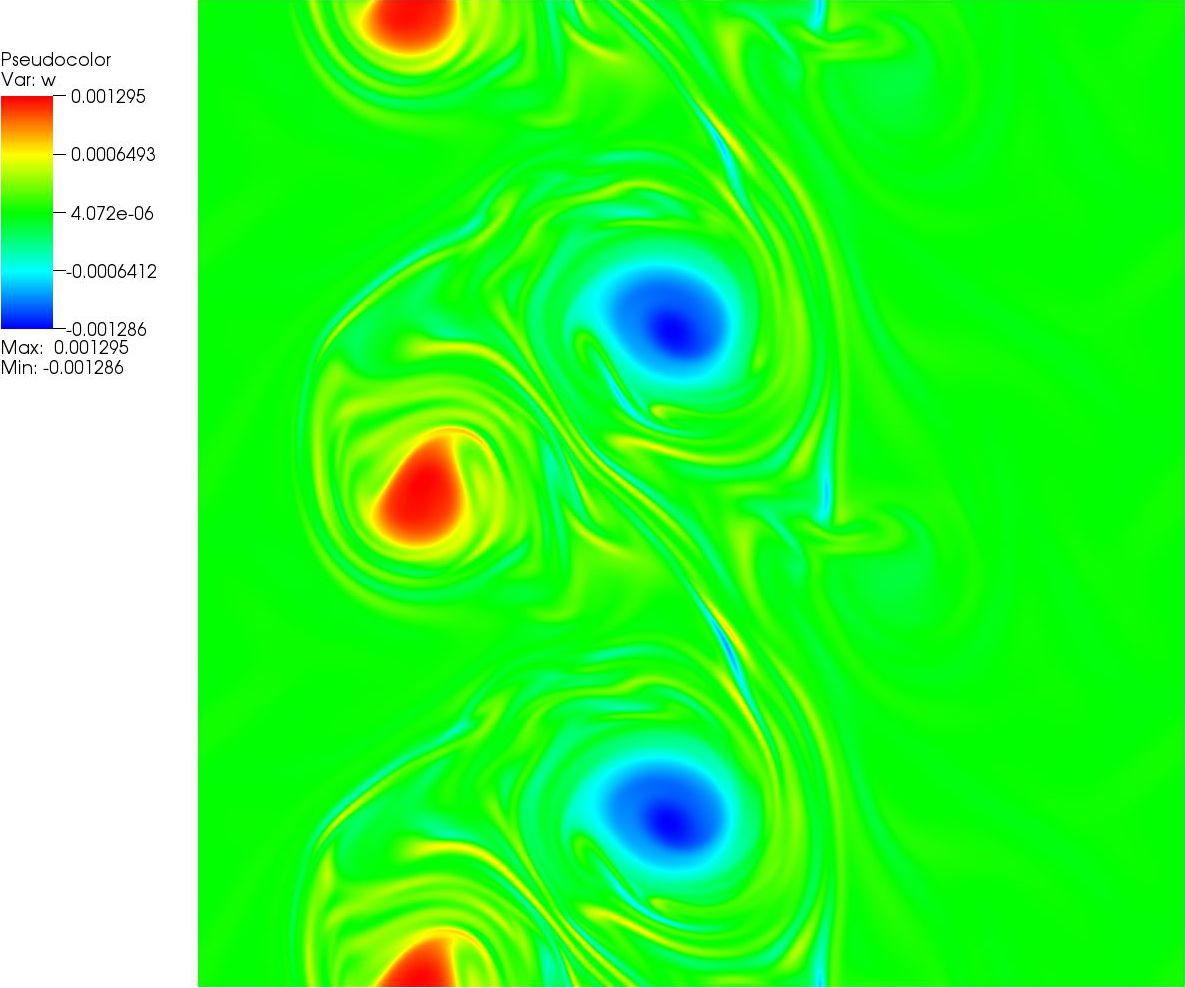}}&
			$t=400$k
			\vspace{.05in}
			\\
			
			\raisebox{-0.5\height}{\includegraphics[width=2in, height=2.3in, keepaspectratio=true]{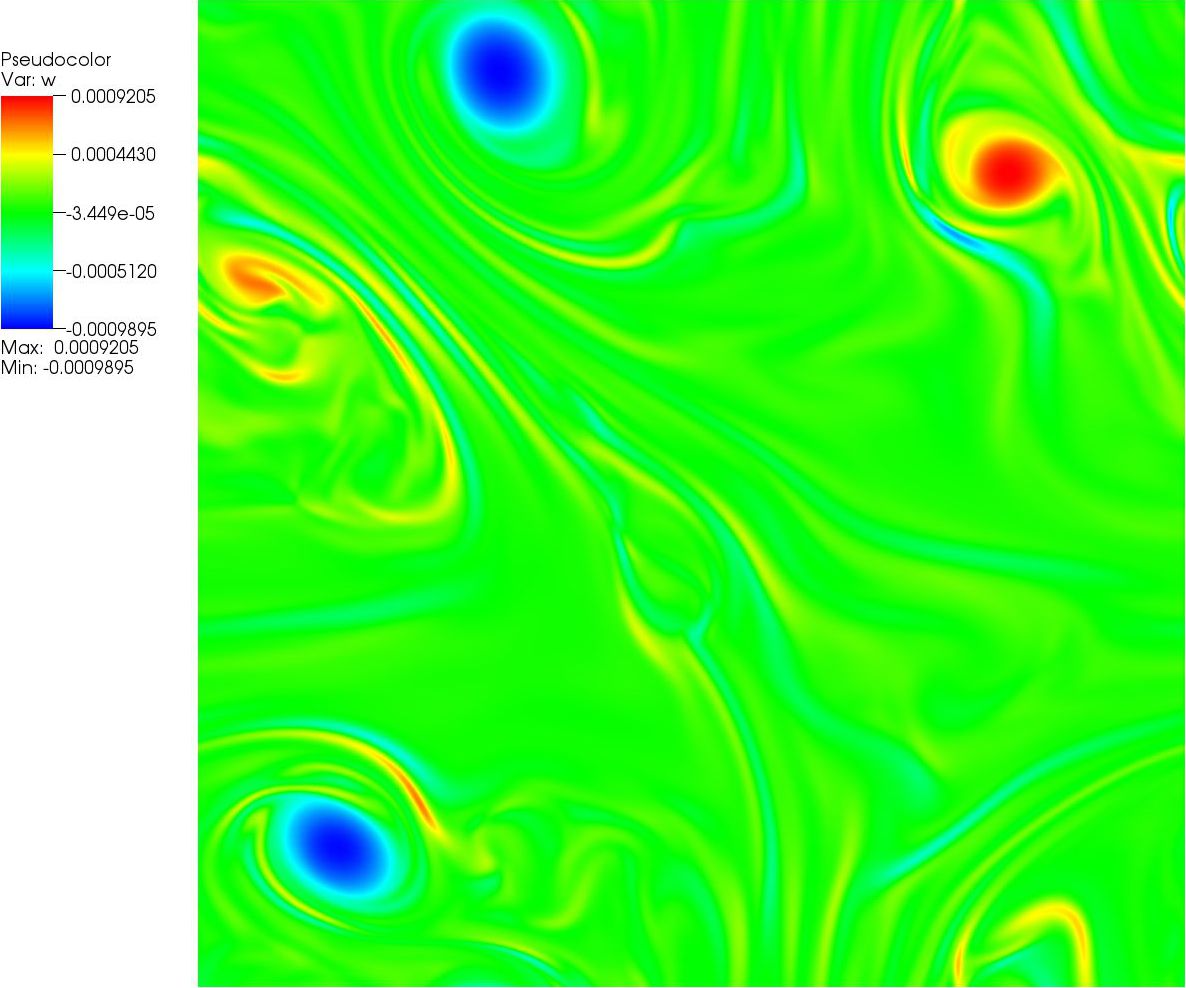}}&
			$t=780$k
		\end{tabular}
		\caption{LES ($1024^2$ $\Delta = 2$)}
	\end{subfigure}%
	\caption{\label{LESEvol} The vorticity evolution of the unstable magnetized Kelvin-Helmholtz jet. \\(a) On the left, the DNS simulation on grid $=2048^2$ ; \\(b) on the right, the LES simulation on grid $=1024^2$ and $\Delta = 2$.\\There is excellent agreement between DNS and LES simulations with time scaling $t_{\mathrm{DNS}} = 2 t_{\mathrm{LES}}$ due to the chosen grids.}
	
\end{figure}

\begin{figure}
	\centering
	\begin{subfigure}[b]{0.4\textwidth}
		\includegraphics[width=2in, height=2.3in, keepaspectratio=true]{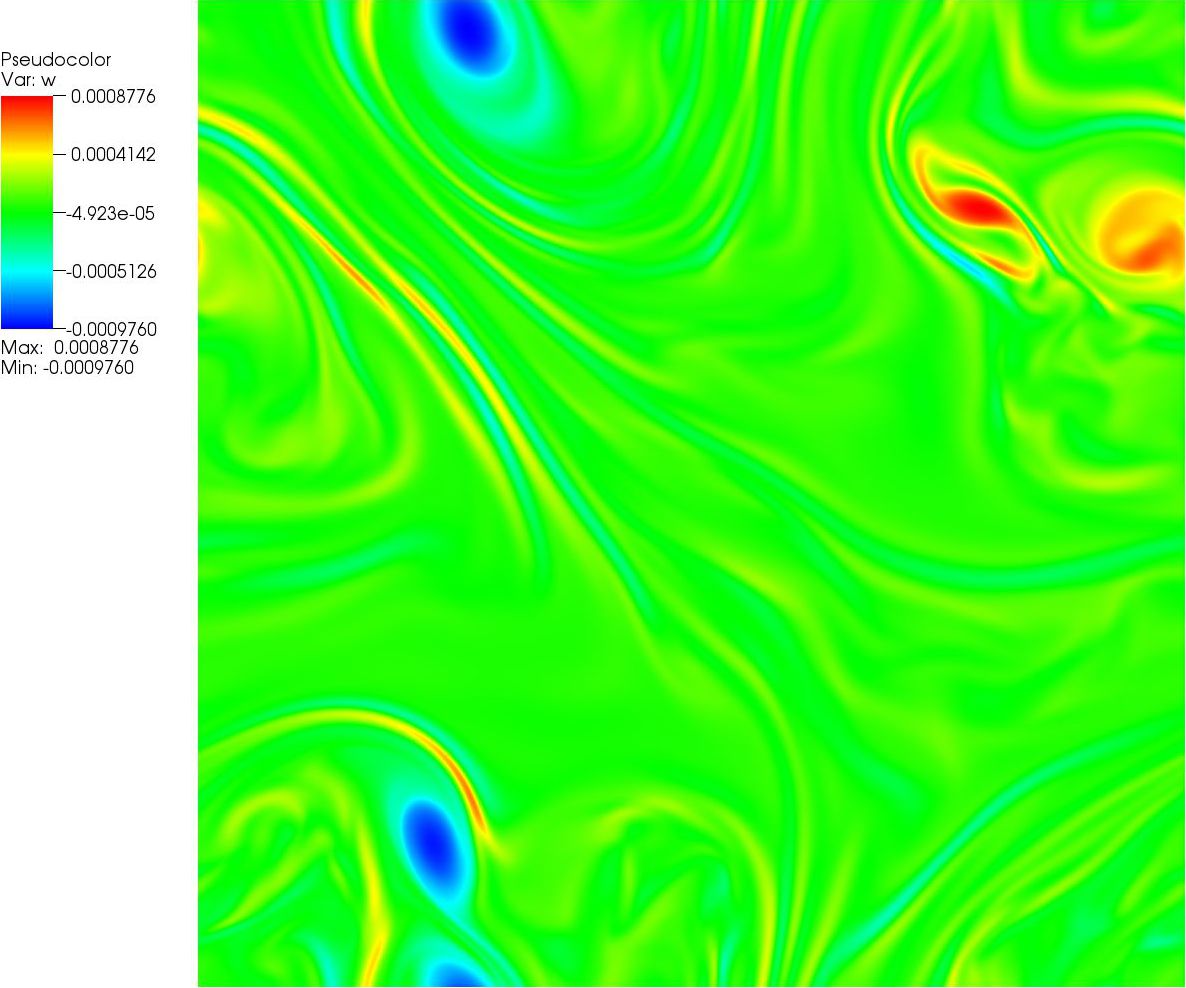}
		\caption{DNS $1024^2$}
	\end{subfigure}%
	\begin{subfigure}[b]{0.4\textwidth}
		\includegraphics[width=2in, height=2.3in, keepaspectratio=true]{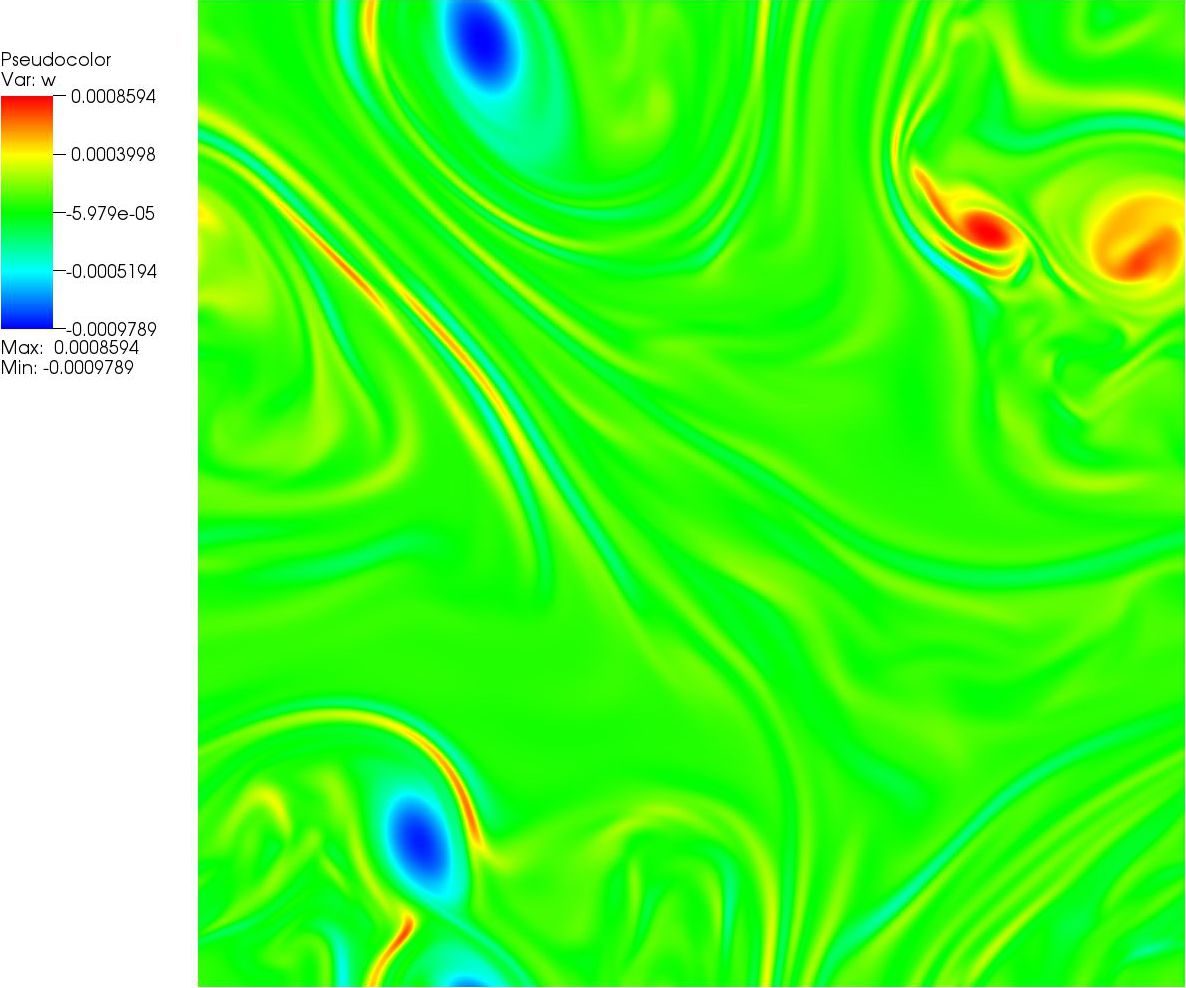}
		\caption{LES $1024^2$, $\Delta = 1$}
	\end{subfigure}
	\begin{subfigure}[b]{0.4\textwidth}
		\includegraphics[width=2in, height=2.3in, keepaspectratio=true]{DNS2w1560k.jpg}
		\caption{DNS $2048^2$}
	\end{subfigure}%
	\begin{subfigure}[b]{0.4\textwidth}
		\includegraphics[width=2in, height=2.3in, keepaspectratio=true]{LES1d4w780k.jpg}
		\caption{LES $1024^2$, $\Delta = 2$}
	\end{subfigure}
	
	\caption{\label{FinalLES} A late time vorticity snapshot comparison between (a) DNS on grid $=1024^2$ at $t = 780k$, \\(b) LES on grid $=1024^2$ with $\Delta = 1$, at $t = 780k$, (c) DNS on grid $=2048^2$ at $t = 1.56M$, and \\(d) LES on grid $=1024^2$ at $t = 780k$ but with filter width $\Delta = 2$.}
\end{figure}

\begin{figure}
	\centering
	\begin{subfigure}[b]{0.4\textwidth}
		\includegraphics[width=2in, height=2.3in, keepaspectratio=true]{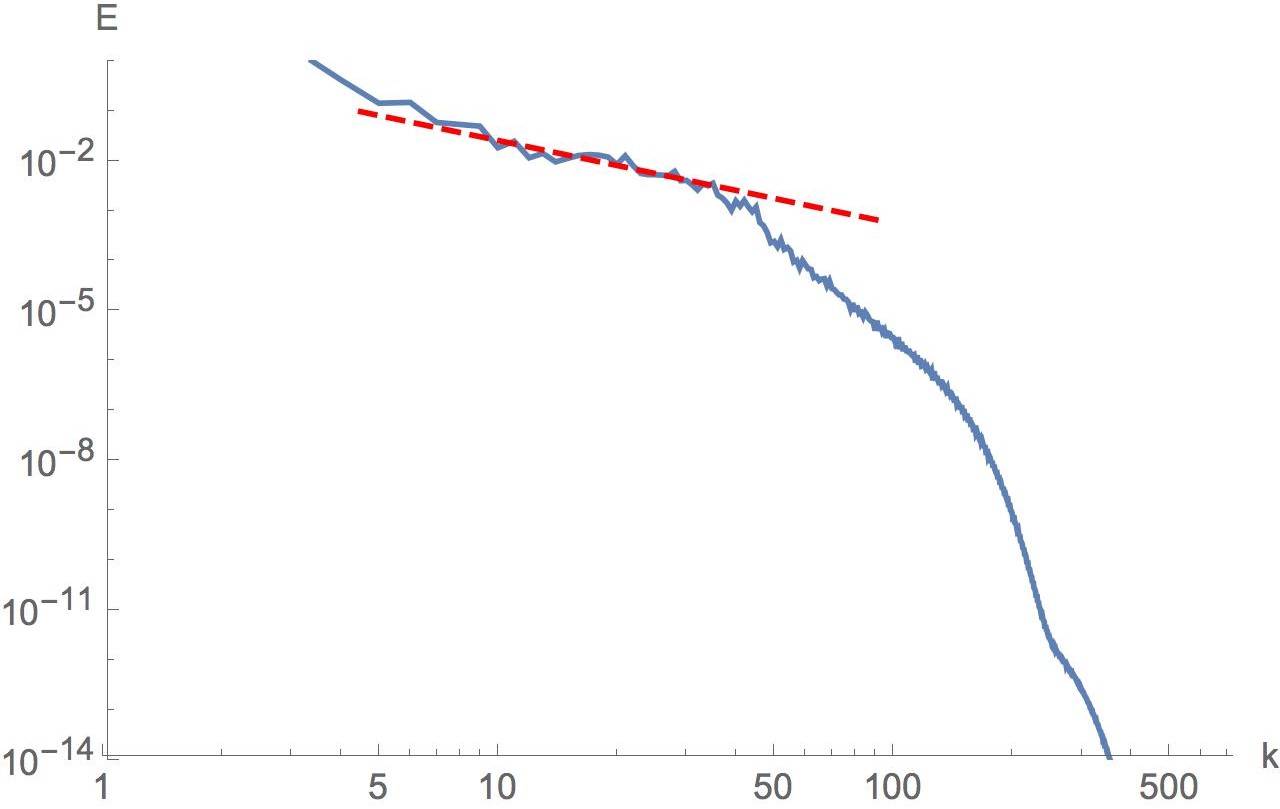}
		\caption{DNS $1024^2$}
	\end{subfigure}%
	\begin{subfigure}[b]{0.4\textwidth}
		\includegraphics[width=2in, height=2.3in, keepaspectratio=true]{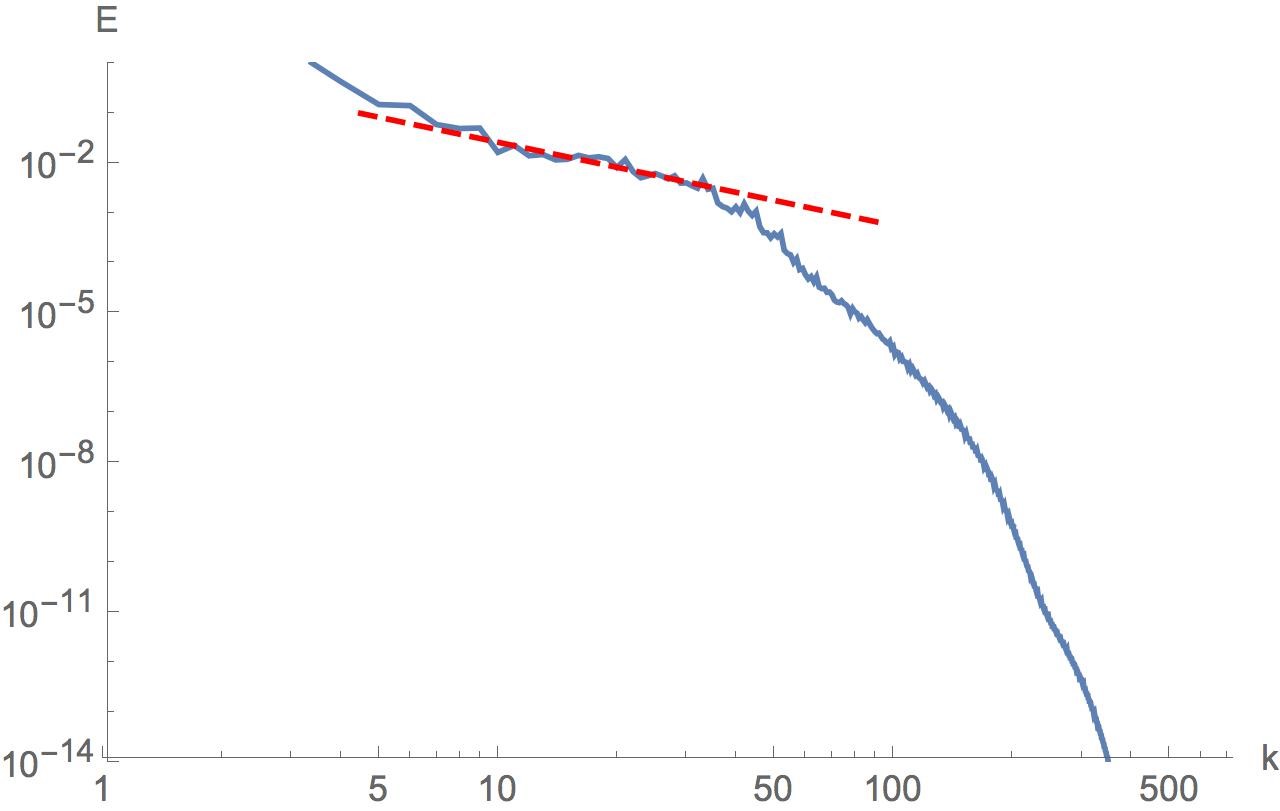}
		\caption{LES $1024^2$, $\Delta = 1$}
	\end{subfigure}
	\begin{subfigure}[b]{0.4\textwidth}
		\includegraphics[width=2in, height=2.3in, keepaspectratio=true]{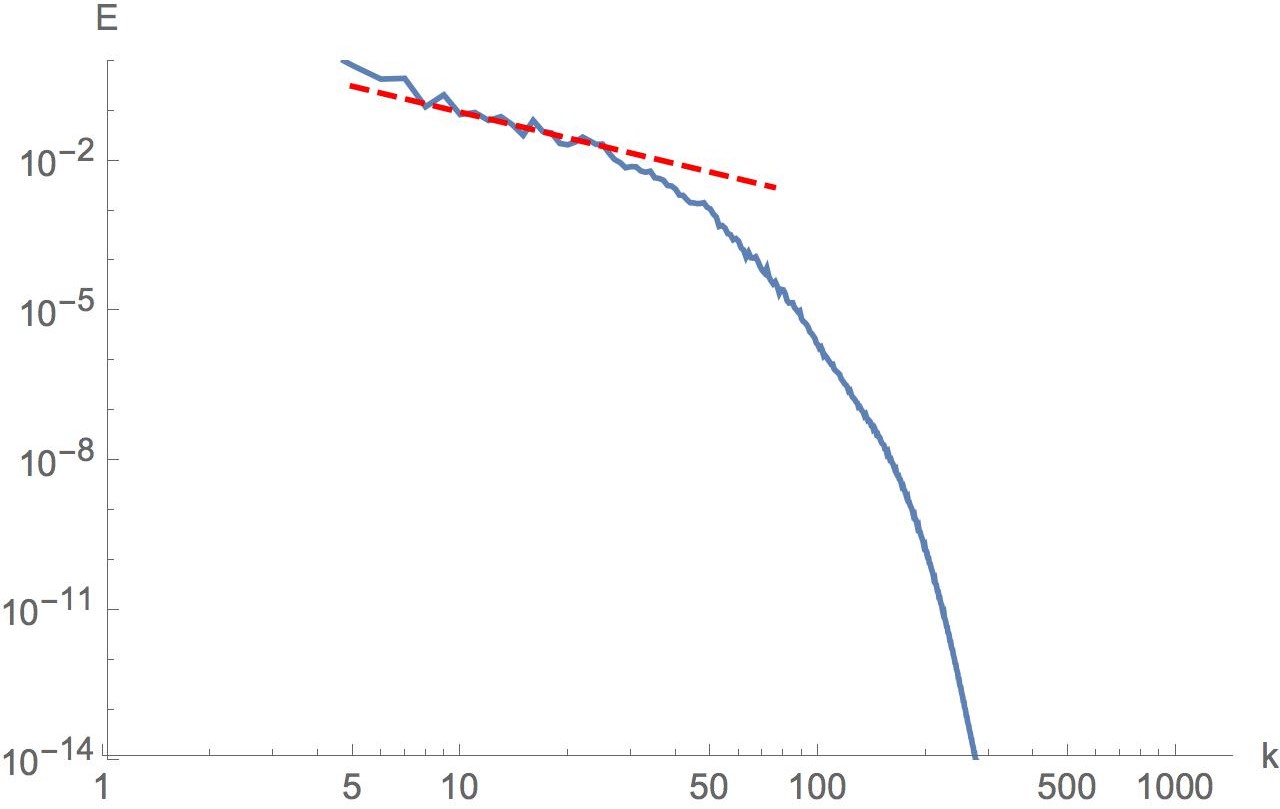}
		\caption{DNS $2048^2$}
	\end{subfigure}%
	\begin{subfigure}[b]{0.4\textwidth}
		\includegraphics[width=2in, height=2.3in, keepaspectratio=true]{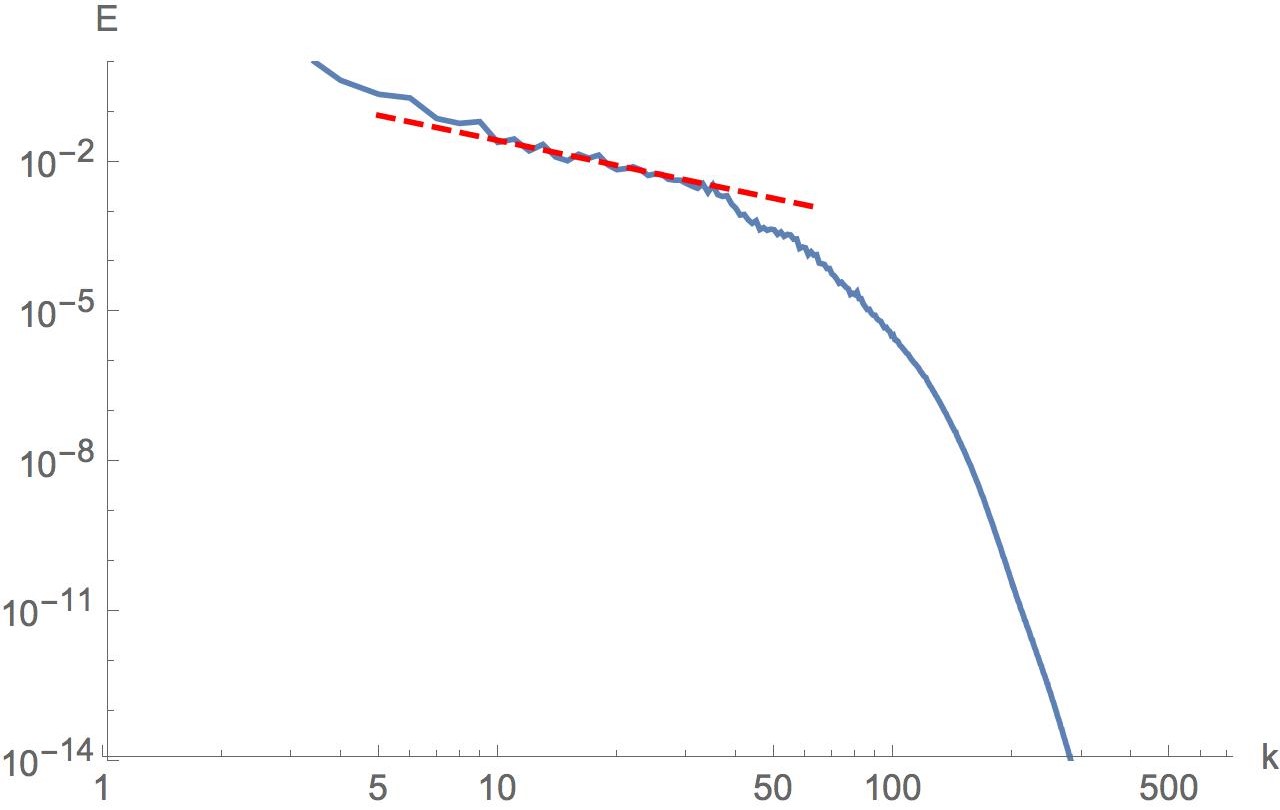}
		\caption{LES $1024^2$, $\Delta = 2$}
	\end{subfigure}
	
	\caption{\label{FinalTotEn} A late time spectral comparison between (a) DNS on grid $=1024^2$ with slope $k^{-1.66}$ at $t = 780k$, (b) LES on grid $=1024^2$ with $\Delta = 1$ and slope $k^{-1.66}$ at $t = 780k$, (c) DNS on grid $=2048^2$ with slope $k^{-1.71}$ at $t = 1.56M$, and (d) LES on grid $=1024^2$ with slope $k^{-1.66}$ at $t = 780k$ but with filter width $\Delta = 2$.}
\end{figure}

\begin{figure}
	\centering
	\begin{subfigure}[b]{0.4\textwidth}
		\includegraphics[width=2in, height=2.3in, keepaspectratio=true]{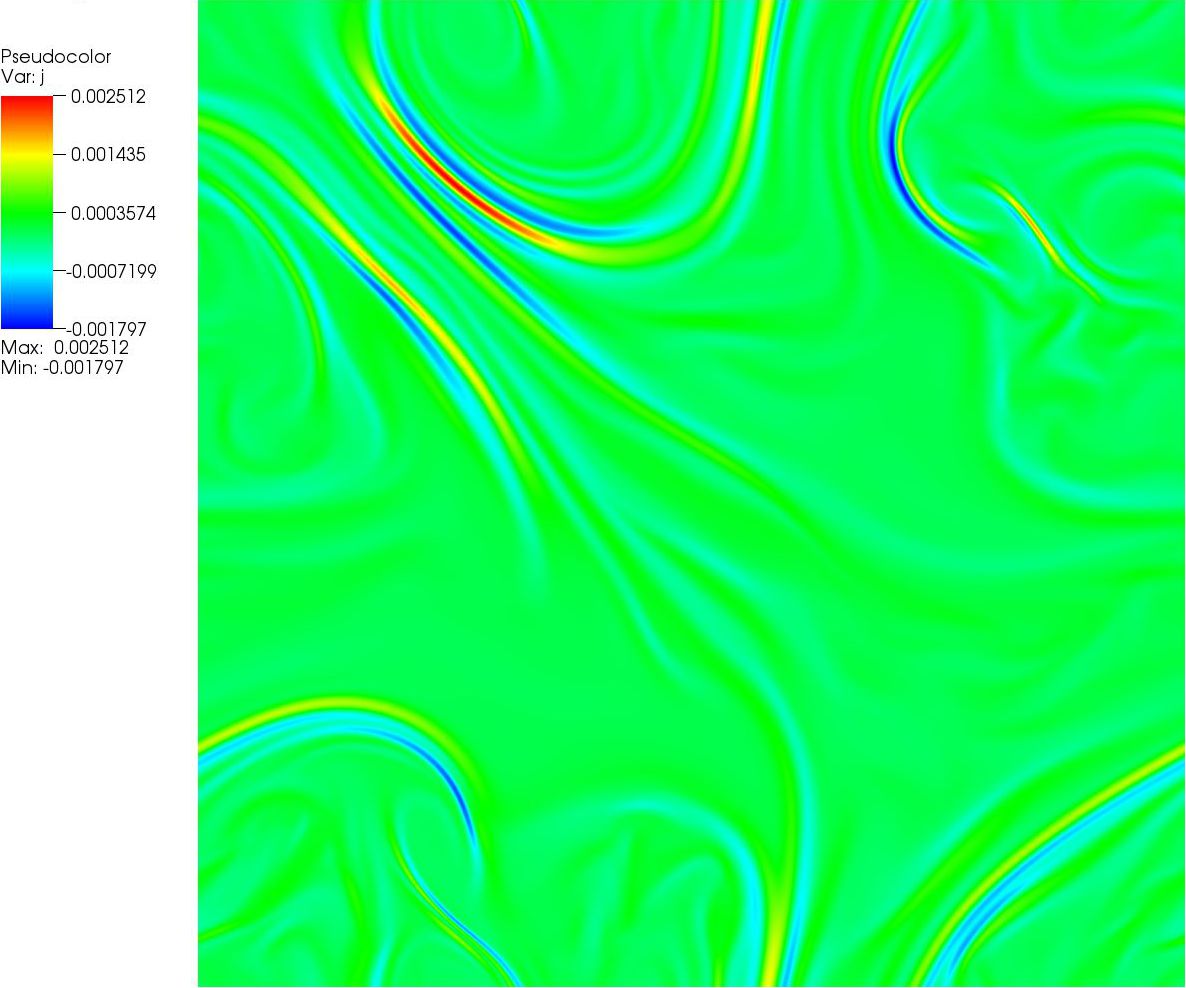}
		\caption{DNS $1024^2$}
	\end{subfigure}%
	\begin{subfigure}[b]{0.4\textwidth}
		\includegraphics[width=2in, height=2.3in, keepaspectratio=true]{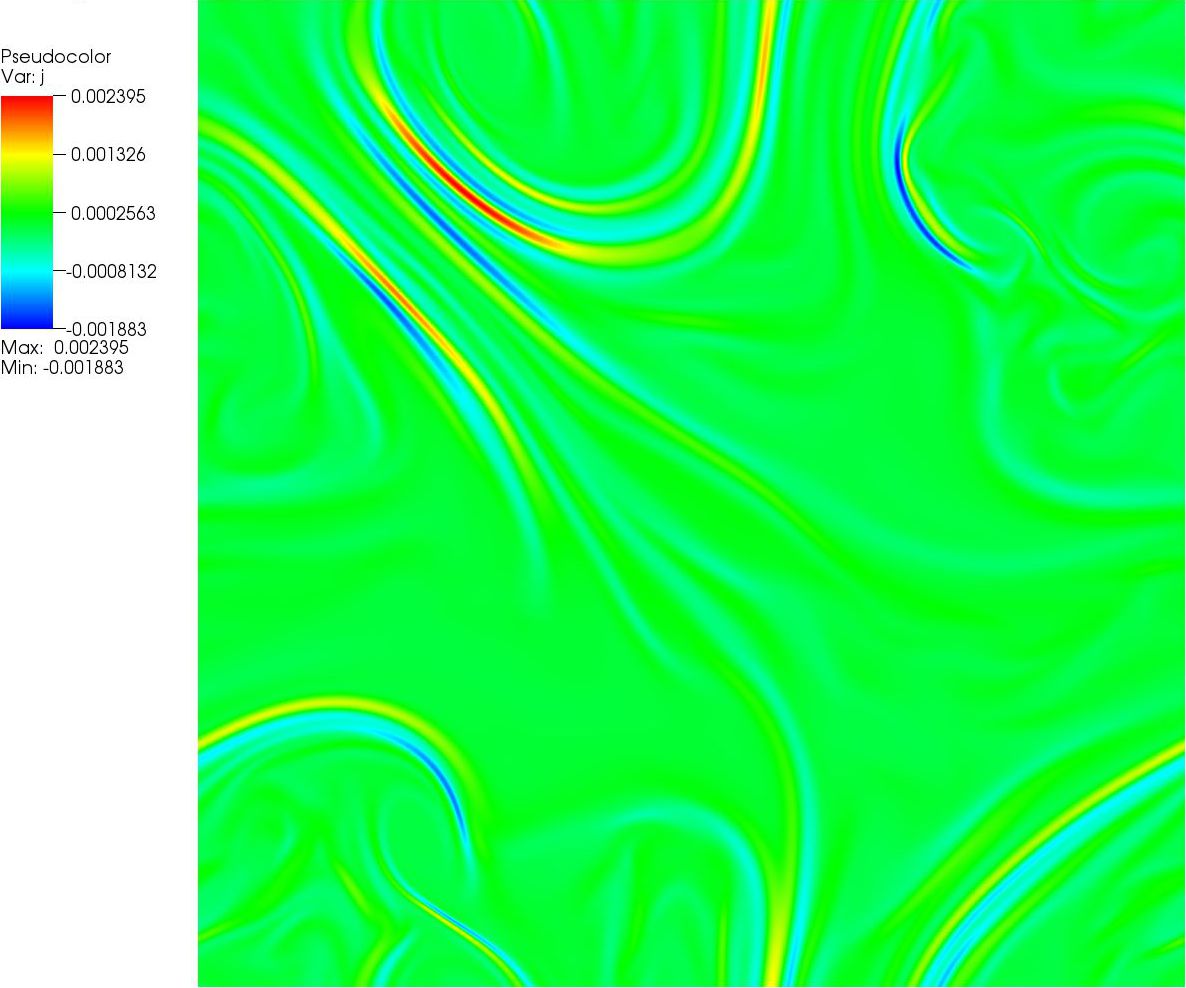}
		\caption{LES $1024^2$, $\Delta = 1$}
	\end{subfigure}
	\begin{subfigure}[b]{0.4\textwidth}
		\includegraphics[width=2in, height=2.3in, keepaspectratio=true]{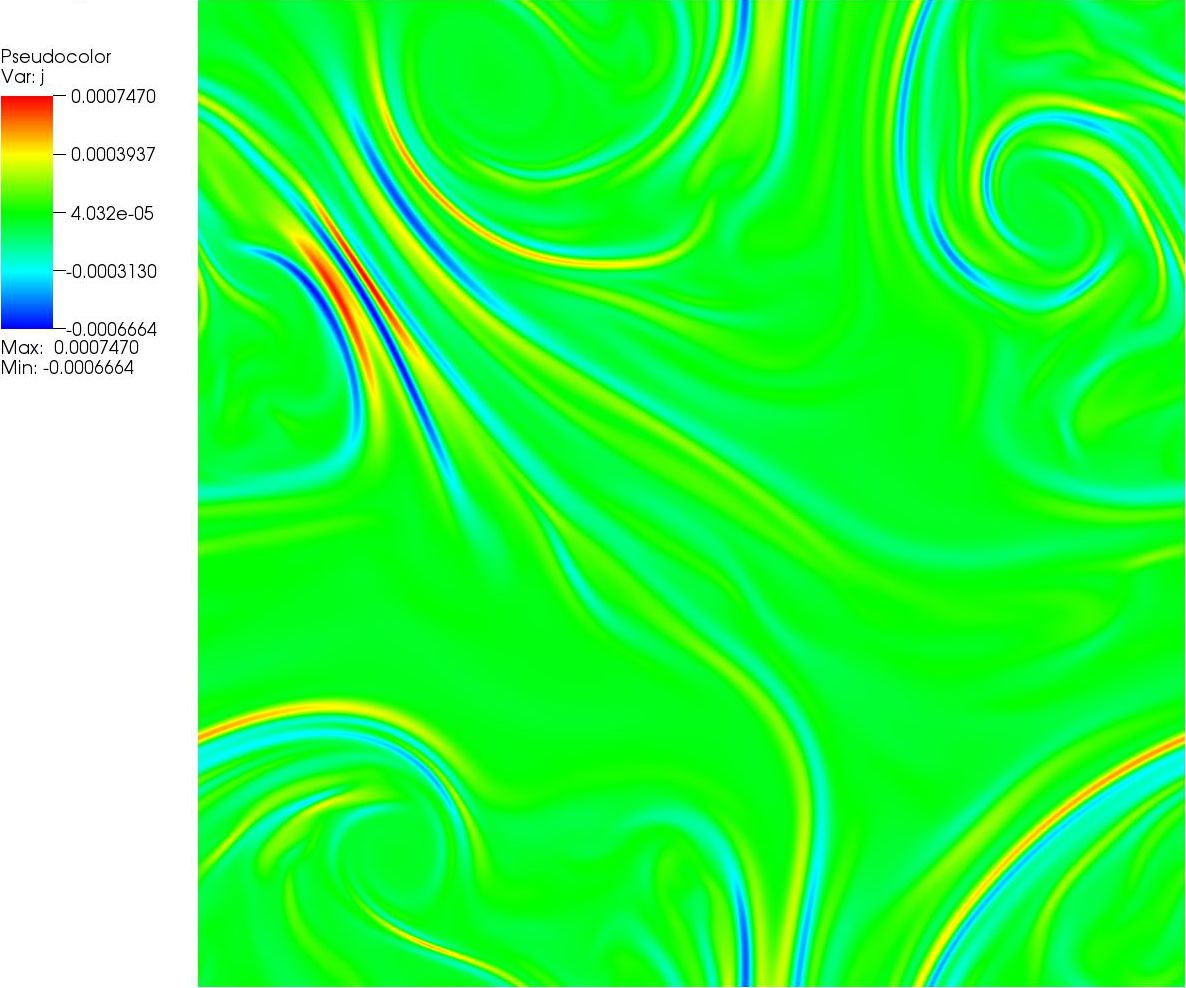}
		\caption{DNS $2048^2$}
	\end{subfigure}%
	\begin{subfigure}[b]{0.4\textwidth}
		\includegraphics[width=2in, height=2.3in, keepaspectratio=true]{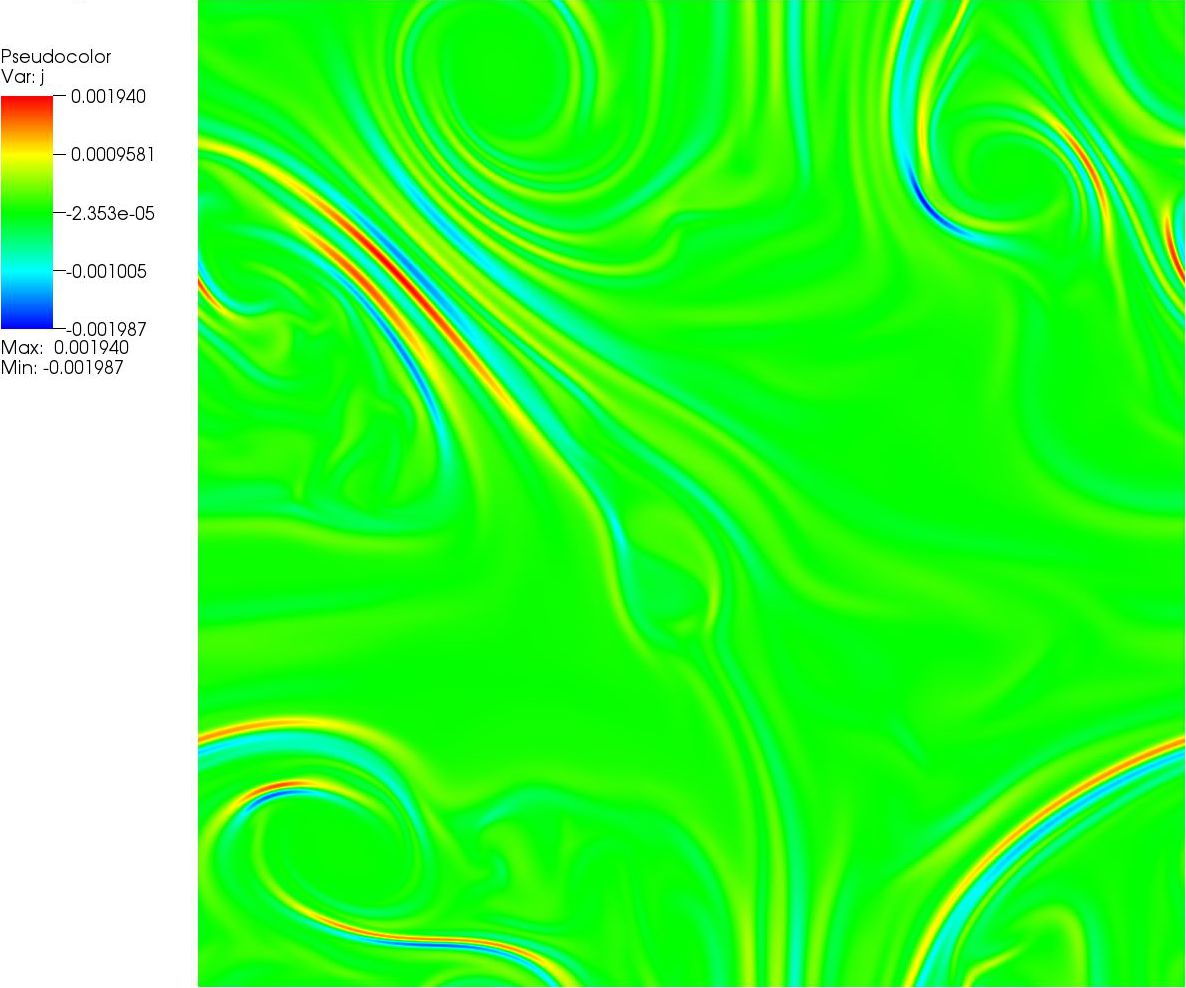}
		\caption{LES $1024^2$, $\Delta = 2$}
	\end{subfigure}
	
	\caption{\label{FinalLESCurrent} A late time snapshot current comparison between (a) DNS on grid $=1024^2$ at $t = 780k$, \\(b) LES on grid $=1024^2$ with $\Delta = 1$, at $t = 780k$, (c) DNS on grid $=2048^2$ at $t = 1.56M$, and \\(d) LES on grid $=1024^2$ at $t = 780k$ but with filter width $\Delta = 2$.}
\end{figure}

It should be noted that as in regular LB-MHD, the filtered $\nabla \cdot \overline{\mathbf B} = 0$ is maintained to machine accuracy.

There is a little subtlety in that not all the spatial derivatives in the filtered collision moments can be determined from local perturbed moments \cite{ChenDoolen,DellarMHDLB}.  This limitation is thought to arise from the low D2Q9 lattice.  It is expected that on a D3Q27 lattice the linearly independent set of derivatives can be represented by the now larger number of local perturbed moments.

While we solve the filtered LB equations, resulting in the filtered LES MHD Eqs. \eqref{FinalLESeq}, there is some similarity in our final MHD model with that of the "tensor diffusivity" model of M{\"u}ller-Carati \cite{Carati2002}.  However it must be stressed that we are performing a first principles derivation of the eddy transport coefficients from a kinetic (LB) model while Muller-Carati propose an ad hoc scheme of minimizing the error between two filters at each time step in their determination of their model's transport coefficients.

We will now evolve in time our filtered LB equations and  consider the magnetized Kelvin-Helmholtz instability in a sufficiently weak magnetic field so that the 2D velocity jet is not stabilized \cite{chandrasekhar}.  The initial jet velocity profile is $U_y = U_0 \sech^2{\left(\frac{2\pi}{L}4x\right)}$.  The corresponding vorticity is shown in Fig. \ref{Initial}.  
The initial Reynolds number is chosen to be Re = $U_0 L/ \nu  = 50$k = const., with $U_0 = 4.88\times10^{-2}$ and $B_0 = 0.005 U_0$. The viscosity and resistivity on a grid of $1024^2$ are $\nu=\eta=10^{-3}$ and scale with the grid to maintain a constant Re and a constant magnetic Reynolds number $U_0 L/ \eta$.  The initial perturbation to the fields are:  $U_y = 0.01 U_0 \sin{\left(\frac{2\pi}{L}4x\right)}$, $B_y = 0.01 B_0 \sin{\left(\frac{2\pi}{L}4x\right)}$, $U_x = 0.01 U_0 \sin{\left(\frac{2\pi}{L}4y\right)}$, and $B_x = 0.01 B_0 \sin{\left(\frac{2\pi}{L}4y\right)}$.  Note that initially $\nabla \cdot \vec B = 0 = \nabla \cdot \vec U$.

In Fig, \ref{LESEvol} we compare the evolution of vorticity in time from DNS on a $2048^2$ grid with that determined from our LES-LB-MHD model on a $1024^2$ grid.  The DNS simulations are determined  by solving the direct unfiltered LB Eqs. \eqref{LBKinEqn} and \eqref{LBMagEqn}.
For constant Reynolds number simulations at different grid sizes, the kinematic viscosity is adjusted appropriately.  Thus on halving the spatial grid, a DNS time step of $2t_{0}$ corresponds to time step $t_0$ in LES-LB-MHD.

At relatively early times the jet profile width slightly widens while within the vorticity layers the Kelvin-Helmholtz instability will break these layers into the familiar vortex street (Fig. \ref{LESEvol}).  Since we have chosen a weak magnetic field insufficient to stabilize the jet, the vorticity streets break apart with like vortex-vortex reconnection (Fig. \ref{LESEvol}).  There is very good agreement between DNS and LES-LB-MHD with filter width $\Delta=2$ (in lattice units) on a grid $L/2$.

%
%
%

Finally, we show the corresponding vorticity (Fig. \ref{FinalLES}), total energy spectrum (Fig. \ref{FinalTotEn}), and current (Fig. \ref{FinalLESCurrent}) plots at t = 780k for simulations on $1024^2$ grids  and their counterparts on $2048^2$ grids at time t = 1.56M.  
We consider 4 cases: (a) DNS on $1024^2$, (b) filtered LB-LES-MHD on $1024^2$ grid and small filter width, $\Delta = 1$, (c)  DNS  on $2048^2$ and (d) filtered LES-LB-MHD on a $1024^2$ grid but with filter width $\Delta=2$.  The effect of the filter width $\Delta$ in our LB-LES-MHD model on the evolution of the vorticity is evident when comparing  Fig. \ref{FinalLES}b to Fig. \ref{FinalLES}d - both in location and strength of the main vortices as well as in the fine grained small scale vorticity.  
As the filter width increases to $\Delta = 2$ (fig. \ref{FinalLES}d), there is stronger agreement now with the DNS (fig. \ref{FinalLES}c) on $L^2$ grid with our LES-LB-MHD filtered model on $(L/2)^2$ grid.  This shows that the subgrid terms are now influencing larger scales with some accuracy. 
The spectral plots (fig. \ref{FinalTotEn}) are somewhat similar in all simulations with a very localized Kolmogorov energy spectrum.  Presumably this is because the turbulence is limited and relatively weak.  There appears to be good agreement in both the vorticity and current between DNS and LES-LB-MHD with $\Delta=2$ on half the grid.


\section{Conclusion}

Here we have presented some preliminary 2D filtered SRT LB-MHD simulation results based on an extension of ideas of Ansumali \textit{et. al.} \cite{Ansumali} that leads to a self-consistent LES-LB closure scheme based solely on expansions in the filter width $\Delta$ and invoking the constraint that any eddy transport effects can only occur on the transport time scales.  We find very good agreement between DNS and our LES-LB-MHD models.   This warrants further investigation of other filters used in LES, as well as in dynamic subgridding commonly used in LES of Navier-Stokes turbulence. Finally, an exploration of the effects of MRT on this LES algorithm should be quite interesting as a somewhat unexpected term related to the gradient of a pressure appears in the subgrid viscosity. This term reveals that higher-order moments (not stress related) can have a first order effect on the subgrid viscosity when MRT is employed. Given that this subgrid pressure term relies on the existence of higher order moments, it suggests that the extra parameters in lattice Boltzmann (ie. the distribution velocities/moments) are introducing new physics naturally absent from LES in computational fluid dynamics. It would be very interesting to see whether this new term enhances the LES accuracy or increases stability at even higher Reynold's flow. Further study could include how this term effects other, well-established LES approaches in computational fluid dynamics. These ideas are under consideration.

\section{Acknowledgments}
This work was partially supported by an AFOSR and NSF grant.  The computations were performed on Department of Defense supercomputers.

\section*{References}
\bibliography{LESPaperBib}
   
 \end{document}